\documentclass[11pt, twocolumn]{article}

\usepackage{times}
\usepackage{latexsym}
\usepackage{microtype}
\usepackage{natbib}
\usepackage{geometry}
\usepackage{xurl} 

\usepackage{caption}
\usepackage{times}
\usepackage{latexsym}

\usepackage[T1]{fontenc}

\usepackage[utf8]{inputenc}

\usepackage{microtype}

\usepackage{inconsolata}

\usepackage{graphicx}
\usepackage{hyperref}
\hypersetup{
    colorlinks=true,
    linkcolor=blue,
    citecolor=blue,
    filecolor=magenta,
    urlcolor=blue,
}

\usepackage{adjustbox}
\usepackage{graphicx}
\usepackage{tikz}
\usepackage{enumitem}
\usepackage{forest}
\usetikzlibrary{trees,positioning,shapes,shadows,arrows.meta}

\usepackage{booktabs}
\usepackage{array}
\usepackage{tabularx}
\usepackage{hyperref}
\usepackage{longtable}
\usepackage{supertabular}
\usepackage{rotating}   
\usepackage{ragged2e}

\usepackage{xcolor}
\usepackage{pifont}
\usepackage{multirow}
\usepackage{colortbl}
\definecolor{headerblue}{RGB}{220, 225, 240}
\usepackage{ragged2e}

\usepackage{calc}

\newcolumntype{L}[1]{>{\RaggedRight\arraybackslash}p{#1}}
\newcolumntype{Y}{>{\RaggedRight\arraybackslash}X}

\title{LLM Agents in Law: Taxonomy, Applications, and Challenges}

\author{
 \textbf{Shuang Liu\textsuperscript{1}},
 \textbf{Ruijia Zhang\textsuperscript{2}},
 \textbf{Ruoyun Ma\textsuperscript{3}},
 \textbf{Yujia Deng\textsuperscript{4}},
\\
 \textbf{Lanyi Zhu\textsuperscript{5}},
 \textbf{Jiayu Li\textsuperscript{6}},
 \textbf{Zelong Li \textsuperscript{7}},
 \textbf{Zhibin Shen \textsuperscript{8}},
 \textbf{Mengnan Du \textsuperscript{9}}
\\
\\
 \textsuperscript{1}Carnegie Mellon University,
 \textsuperscript{2}National University of Singapore,
 \textsuperscript{3}Stanford University,
\\
 \textsuperscript{4}Independent Researcher,
 \textsuperscript{5}University of Washington,
 \textsuperscript{6}The University of Chicago,
\\
 \textsuperscript{7}Rutgers University,
 \textsuperscript{8}Columbia University,
 \textsuperscript{9}New Jersey Institute of Technology
\\
}

\date{}

\begin{document}
\maketitle

\begin{abstract}
Large language models (LLMs) have precipitated a dramatic improvement in the legal domain, yet the deployment of standalone models faces significant limitations regarding hallucination, outdated information, and verifiability. Recently, LLM agents have attracted significant attention as a solution to these challenges, utilizing advanced capabilities such as planning, memory, and tool usage to meet the rigorous standards of legal practice. In this paper, we present a comprehensive survey of LLM agents for legal tasks, analyzing how these architectures bridge the gap between technical capabilities and domain-specific needs. Our major contributions include: (1) systematically analyzing the technical transition from standard legal LLMs to legal agents; (2) presenting a structured taxonomy of current agent applications across distinct legal practice areas; (3) discussing evaluation methodologies specifically for agentic performance in law; and (4) identifying open challenges and outlining future directions for developing robust and autonomous legal assistants.
\end{abstract}

\section{Introduction}

In recent years, large language models (LLMs) have precipitated a dramatic improvement in the legal domain, fundamentally altering how legal professionals approach complex information processing. These models have demonstrated exceptional capability across a variety of specialized legal tasks, ranging from routine document processing to sophisticated reasoning challenges. The integration of LLMs has improved distinct application areas, including legal judgment prediction~\cite{shu2024lawllm}, legal question answering~\cite{louis2024interpretable}, and contract review~\cite{liu2025contracteval}, offering new efficiencies in workflows that were previously manual and labor-intensive.

However, the deployment of standalone LLMs for legal tasks faces significant limitations, primarily due to persistent issues such as hallucination~\cite{farquhar2024detecting,sriramanan2024llm} and the generation of outdated information. Given that legal practice constitutes a high-stakes environment where precision is mandatory and errors can have severe consequences, these reliability issues severely restrict the application of standard LLMs in real-world scenarios. To address these critical challenges, the field is shifting toward the development of LLM Agents~\cite{li2025legalagentbench}, which utilize advanced capabilities such as planning, memory, and tool usage, to mitigate the shortcomings of base LLM models and meet the rigorous standards required by the legal profession.

In this paper, we present a comprehensive survey of LLM Agents for legal tasks, structuring our analysis to bridge the gap between technical capabilities and domain-specific needs. We first examine why LLM Agents are particularly promising for this domain by contrasting their advantages against the limitations of standalone models, followed by a detailed review of current application areas. Furthermore, we assess the performance of existing agents and conclude by highlighting open challenges and future directions to guide the next phase of research in legal AI.

\subsection{Contribution and Uniqueness}
\noindent\textbf{Our Contributions.}
This paper provides a comprehensive overview of LLM agents in the legal domain, with four major contributions: (1) We systematically analyze the technical transition from standard legal LLMs to legal agents, detailing how agentic characteristics address critical deficiencies in legal applications. (2) We present a structured taxonomy of current agent applications across five distinct legal practice areas. (3) We discuss evaluation methodologies specifically for agentic performance in law. (4) We identify open challenges and outline future directions for developing robust, autonomous legal assistants.

\vspace{3pt}
\noindent\textbf{Differences with Existing Surveys.}
While several existing surveys examine AI application in law~\cite{lai2024large,hou2025large,chen2024survey}, they differ significantly in scope and focus compared to our agent-centric approach. For instance, Lai et al.~\cite{lai2024large} focuses on standalone model capabilities and general ethical challenges in the judicial system, largely overlooking agentic architectures that utilize planning and tools to address identified limitations. In contrast, our survey uniquely and exclusively focuses on the depth of LLM Agents in the legal field, providing a detailed analysis of agentic principles, workflows, and evaluation criteria specific to legal practice.

\section{Why LLM Agents for Legal Tasks?}  

\begin{figure*}[!ht]
    \centering
    \includegraphics[width=\textwidth]{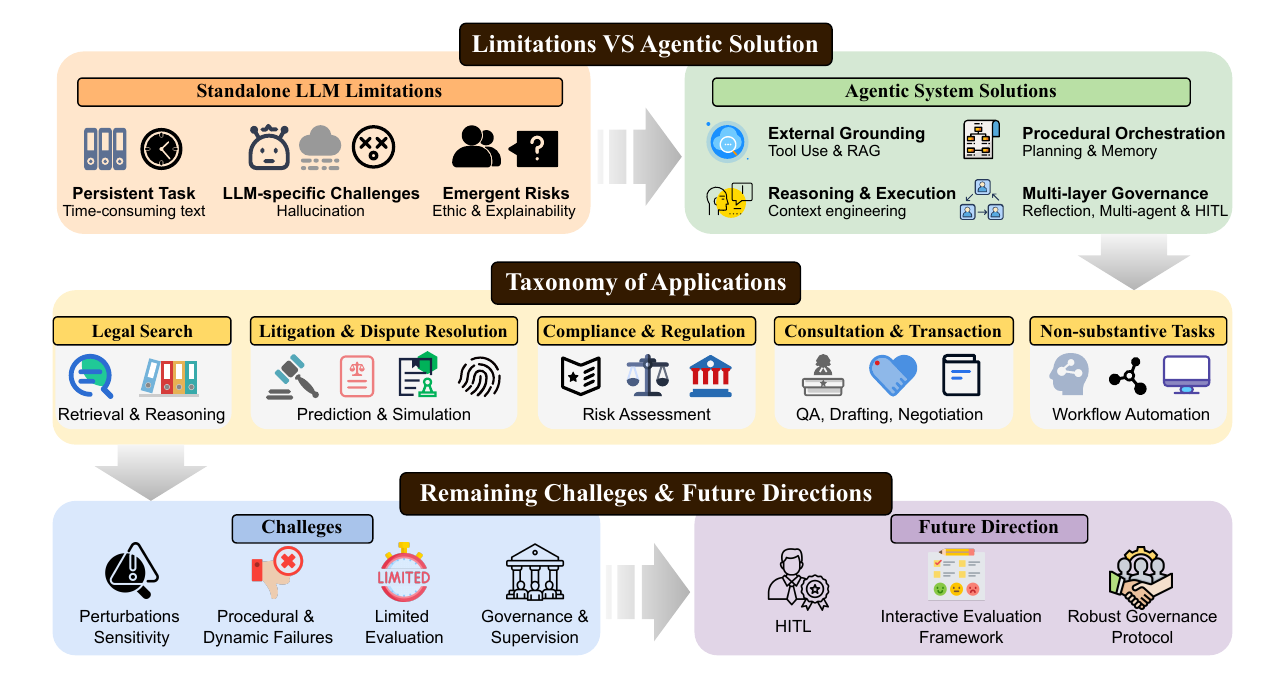}
    \caption{Overview of paper structure and LLM agent application.}
    \label{fig:fig1}
\end{figure*}

\subsection{Why LLMs by themselves are insufficient}

LLMs are reshaping legal workflows through the Transformer architecture's dynamic semantic capture and generative reasoning \cite{shao2025WhenLarge}. They excel at fundamental legal tasks \cite{davenport2025EnhancingLegal}, with further improvements achieved via judicial-syllogism-based Chain-of-Thought (CoT) \cite{song2025LegalText}. Models like LawLLM \cite{shu2024lawllm} demonstrates promise in simulating professional legal reasoning, while newest multimodal models extends analysis to audio and video modalities \cite{VLexTransforming}. However, LLMs do not resolve all legal challenges and even introduce new risks \cite{lai2024large, dehghani2025LargeLanguage}, which we categorize into three classes:
\begin{itemize}[leftmargin=*]\setlength\itemsep{-0.3em}
\item 
\emph{Class A: Persistent Traditional Challenges.} Long-cycle, multi-stage workflows remain difficult for LLMs to navigate~\cite{mohsin2025FundamentalLimits}. Current LLMs struggle with extended task consistency and procedural depth~ \cite{fei-etal-2024-lawbench, liu-etal-2024-lost} for legal tasks requiring complex multi-steps~\cite{guhaLegalBenchCollaborativelyBuilt2023}.

\item \emph{Class B: LLM-Specific Challenges.} Hallucination is inherent in current generative architectures~\cite{xuHallucinationInevitableInnate2025, kalaiWhyLanguageModels2025, huangSurveyHallucinationLarge2025}. In law, fabricated citations or documents pose severe risks and significant real-world consequences~\cite{dahl2024LargeLegal,magesh2025hallucination,2023CaseAI}.

\item \emph{Class C: Exacerbated or Emergent Risks.} Fixed post-training weights hinder adaption to evolving regulations\cite{ferraris2025ArchitectureLanguage}, creating hidden risks where erroneous conclusions are hard to detect or trace\cite{dahl2024LargeLegal}. Further, the ``black-box'' nature of LLMs complicates legal accountability, ethical frameworks, and societal trust\cite{CoreViews, weidinger2022TaxonomyRisks}.
\end{itemize}

\subsection{LLM Agents Address These Limitations}

Although comprehensive lifecycle management best supports responsible LLM development~\cite{wang2025SurveyResponsible}, it is often cost-prohibitive, prompting a shift toward agentic frameworks as a practical alternative to standalone models. Unlike standalone models, LLM agents serve as reasoning engines that orchestrate external modules and iterative workflows \cite{BuildingEffective, xi2023RisePotentiala}. We summarize the core agentic capabilities and corresponding remedies as follows:
\begin{itemize}[leftmargin=*]\setlength\itemsep{-0.3em}
\item \emph{External Grounding and Knowledge Freshness.} To mitigate hallucinations (Class B) and knowledge expiration (Class C), agents employ Tool Use and Retrieval-Augmented Generation (RAG)~\cite{weng2023LLMPowered, lewisRetrievalAugmentedGenerationKnowledgeIntensive2021}. By offloading ``authority'' from internal parameters to verifiable systems such as legal databases, statutes, and case law APIs, gents anchor conclusion to explicit evidence~\cite{cui2023chatlaw, guhaLegalBenchCollaborativelyBuilt2023, huang2023LawyerLLaMA}. This is essential for cross-jurisdictional tasks and rapidly evolving regulatory updates that static models fail to reliably trace~\cite{vu2023FreshLLMsRefreshing}.

\item 
\emph{Procedural Orchestration and Long-term Consistency.} To address persistent traditional challenges (Class A), agents employ Planning and Memory modules. Legal workflows are inherently multi-stage and long-cycle~\cite{li2025legalagentbench, SingleagentMultiagent}, and planning decomposes complex tasks, such as due diligence or multi-step litigation strategy, into manageable sub-goals~\cite{chen2025EnhancingLLMBased}. Meanwhile, memory preserves context over time, preventing the ``lost-in-the-middle'' phenomenon and maintaining consistent legal reasoning across a case's lifecycle~\cite{liu-etal-2024-lost, cui2023chatlaw}.

\item 
\emph{Multi-layer Verification and Governance.} To tackle accountability and ethical risks (Class C), agentic frameworks introduce Reflection, Multi-agent Collaboration, and Human-in-the-loop (HITL) protocols \cite{orbanNewEra, xi2023RisePotentiala}. 
Reflection enables post-generation consistency checks to detect logical contradictions or evidence gaps, effectively mitigating bias \cite{shinn2023ReflexionLanguage, zhang2025mitigating}. 
Multi-agent Systems adopt specialized agents to cross-examine outputs, mimicking the peer-review process in law firms to enhance interpretability \cite{jiangAgentsBench2025, sun2024lawluo, jing2025maslegalbench}. 
HITL positions human as the final arbiter, ensuring data privacy control and ultimate legal accountability\cite{school2025RethinkingHuman, bommarito2025GoverningAI}.
\end{itemize}

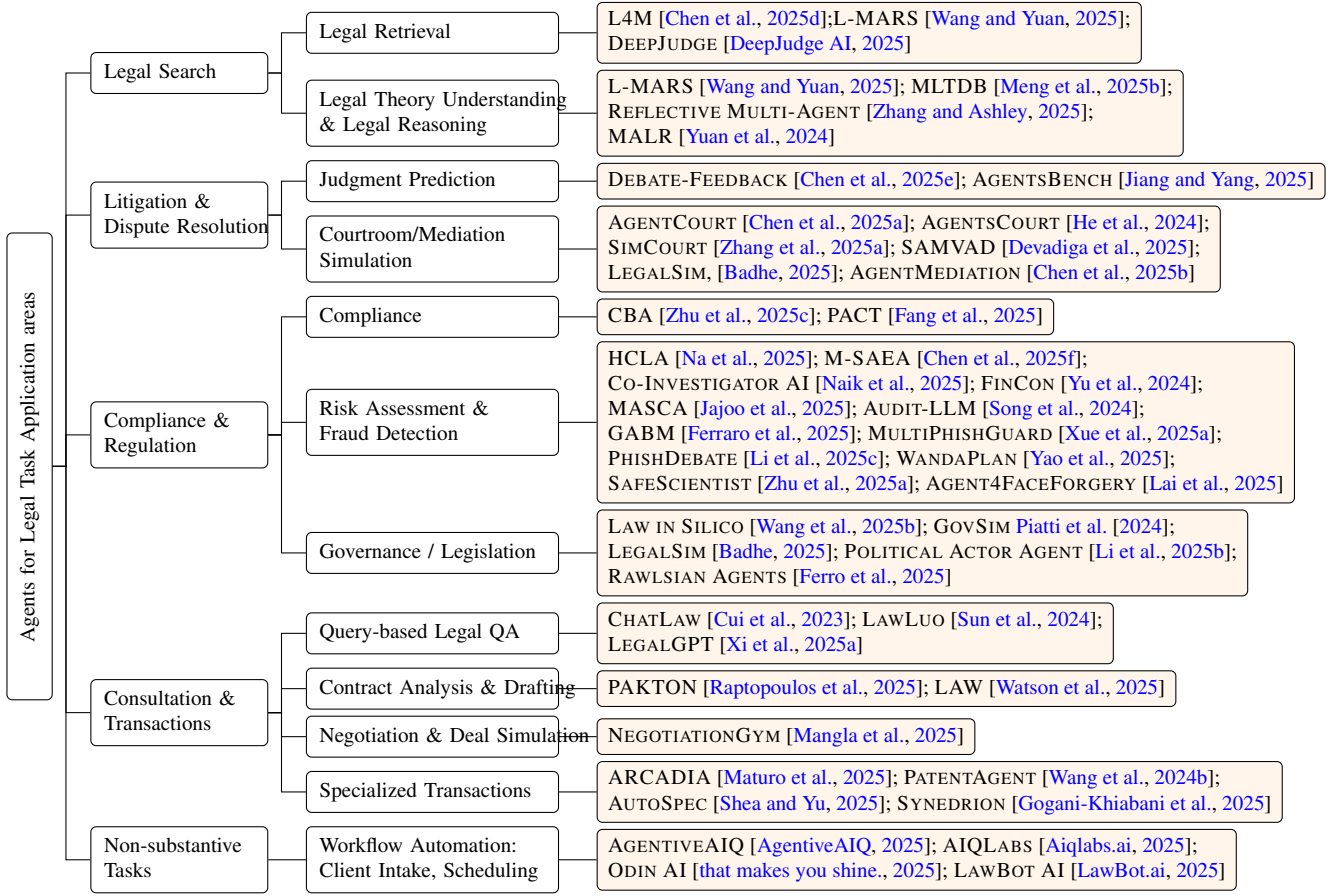
\begin{figure*}[!ht]
\centering
\scriptsize

\forestset{
  boxnode/.style={
    draw=black,
    thin,
    rounded corners=2pt,
    fill=white,
    align=left,
    inner xsep=5pt,
    inner ysep=3pt,
    font=\scriptsize
  },
  citenode/.style={
    draw=black,
    thin,
    rounded corners=2pt,
    fill=orange!8,
    align=left,
    inner xsep=4pt,
    inner ysep=2pt,
    font=\scriptsize
  }
}

\begin{adjustbox}{max width=\textwidth}
\begin{forest}
for tree={
  grow=east,
  reversed=true,
  growth parent anchor=east,
  parent anchor=east,
  child anchor=west,
  anchor=west,
  boxnode,
  edge path={
    \noexpand\path [draw, thin]
    (!u.parent anchor) -- ++(5pt,0) |- (.child anchor);
  },
  l sep=5mm,
  s sep=0.8mm,
}
%
[{\rotatebox[origin=c]{90}{\parbox{6cm}{\centering Agents for Legal Task Application areas}}}, text width=0.25cm
    %
    [{Legal Search}, text width=2cm
      [{Legal Retrieval}, text width=3.0cm
        [{\textsc{L4M} \citep{chenTrustworthyLegalAI2025};\textsc{L-MARS} \citep{wang2025mars}; \\ \textsc{DeepJudge} \citep{Deepjudge} }, citenode]
      ]
      [{Legal Theory Understanding \\ \& Legal Reasoning}, text width=3.0cm
        [{\textsc{L-MARS} \citep{wang2025mars}; \textsc{MLTDB} \citep{meng2025building};\\ \textsc{Reflective Multi-Agent} \citep{zhang2025mitigating}; \\ \textsc{MALR} \citep{yuan2024can}}, citenode]
table      ]
    ]
    %
    [{Litigation \&\\Dispute Resolution}, text width=2cm
      [{Judgment Prediction}, text width=3.0cm
        [{\textsc{Debate-Feedback} \citep{chen2025debate}; \textsc{AgentsBench} \citep{jiangAgentsBench2025}}, citenode]
      ]
      [{Courtroom/Mediation \\ Simulation}, text width=3.0cm
        [{\textsc{AgentCourt} \citep{chenAgentCourtSimulatingCourt2025a}; \textsc{AgentsCourt} \citep{he-etal-2024-agentscourt}; \\ \textsc{SimCourt} \citep{zhang2025chinese}; \textsc{SAMVAD} \citep{devadiga2025samvad}; \\\textsc{LegalSim}, \citep{badhe2025legalsim};  \textsc{AgentMediation} \citep{chen2025simulating}}, citenode]
      ]
    ]
    %
    [{Compliance \&\\Regulation}, text width=2cm
      [{Compliance}, text width=3.0cm
        [{\textsc{CBA} \citep{zhu2025compliance}; \textsc{PACT} \citep{fang2025privacy}}, citenode]
      ]
      [{Risk Assessment \& \\ Fraud Detection}, text width=3.0cm
        [{\textsc{HCLA} \citep{na2025human}; \textsc{M-SAEA} \citep{chen2025tasks}; \\ \textsc{Co-Investigator AI} \citep{naik2025co}; \textsc{FinCon} \citep{yu2024fincon}; \\ \textsc{MASCA} \citep{jajoo2025masca}; \textsc{Audit-LLM} \citep{song2024audit}; \\ \textsc{GABM} \citep{ferraro2025generative}; \textsc{MultiPhishGuard} \citep{xue2025multiphishguard}; \\ \textsc{PhishDebate} \citep{li2025phishdebate}; \textsc{WandaPlan} \citep{yao2025your}; \\ \textsc{SafeScientist} \citep{zhu2025safescientist}; \textsc{Agent4FaceForgery} \citep{lai2025agent4faceforgery}}, citenode]
      ]
      [{Governance / Legislation}, text width=3.0cm
        [{\textsc{Law in Silico} \citep{wang2025law}; \textsc{GovSim} \citet{piatti2024cooperate}; \\ \textsc{LegalSim} \citep{badhe2025legalsim}; \textsc{Political Actor Agent} \citep{li2025political}; \\ \textsc{Rawlsian Agents} \citep{ferrorawlsian}}, citenode]
      ]
    ]
    %
    [{Consultation \& \\ Transactions}, text width=2cm
      [{Query-based Legal QA}, text width=3.0cm
        [{\textsc{ChatLaw} \citep{cui2023chatlaw}; \textsc{LawLuo} \citep{sun2024lawluo}; \\ \textsc{LegalGPT} \citep{xi2025hybridrag}}, citenode]
      ]
      [{Contract Analysis \& Drafting}, text width=3.0cm
        [{\textsc{PAKTON} \citep{raptopoulos2025pakton}; \textsc{LAW} \citep{watson-etal-2025-law}}, citenode]
      ]
      [{Negotiation \& Deal Simulation}, text width=3.0cm
        [{\textsc{NegotiationGym} \citep{mangla2025negotiationgym}}, citenode]
      ]
      [{Specialized Transactions}, text width=3.0cm
        [{\textsc{ARCADIA} \citep{maturo2025arcadia}; \textsc{PatentAgent} \citep{wang2024patentagent}; \\ \textsc{AutoSpec} \citep{shea2025autospec}; \textsc{Synedrion} \citep{gogani2025tax}}, citenode]
      ]
    ]
    %
    [{Non-substantive \\ Tasks}, text width=2cm
      [{Workflow Automation: \\ Client Intake, Scheduling}, text width=3.0cm
        [{\textsc{AgentiveAIQ} \citep{agentiveaiq2025}; \textsc{AIQLabs} \citep{aiqlabs2025}; \\ \textsc{Odin AI} \citep{odinai2025}; \textsc{LawBot AI} \citep{lawbot2025}}, citenode]
      ]
    ]
]
\end{forest}
\end{adjustbox}

\caption{Taxonomy of LLM agents for legal tasks. This framework categorizes five core legal domains and maps them to representative academic and commercial agentic systems.}
\label{fig:legal_agent_survey_tree}
\end{figure*}

\section{Current LLM Agent Application Areas in the Legal Domain}

In this section, we organize existing work into a taxonomy consisting of five core categories: (1) Legal search, which support the application of legal authorities; (2) Litigation and dispute resolution, where agentic systems enable role specialization and long-horizon interaction; (3) Compliance and regulation, where verification and auditability are essential; (4) Advisory consultation and transactional practice, emphasizing client interaction and document-centric workflows; (5) Non-substantive tasks such as administration automation. Figure \ref{fig:legal_agent_survey_tree} shows the structure of the taxonomy with representative agentic systems. Table \ref{tab:legal_llm_agents} and Table \ref{tab:commercial_legal_ai_agents} summarizes the legal agentic systems proposed by existing literature and commercial products. 

This categorization reflects the structure of real-world legal practice, closely mirroring how work is organized within practice groups of legal institutions. Each category maps to distinct technical demands on agents, ranging from retrieval and reasoning primitives, to long-horizon interaction, procedural compliance, and workflow orchestration.

\subsection{Legal Search \& Research}
Legal retrieval, legal theory understanding, and legal reasoning form the foundation of legal search and research system, jointly supporting how users identify, interpret, and apply legal sources. Legal retrieval connects user information needs to cases, statutes, and regulations, while legal theory understanding captures the doctrinal concepts that govern their interpretation. Legal reasoning integrates these components through structured deduction, analogy, and rule application, supporting legally valid and defensible conclusions.

\paragraph{Legal Retrieval.} Legal retrieval often maps informal queries to authoritative cases, statutes, and regulations \citep{liu2022query}, and must ensure high precision, traceability of retrieved evidence, jurisdictional and temporal correctness, while providing interfaces that support professional inspection and control \citep{liu2023investigating, chenTrustworthyLegalAI2025, wang2025mars}. Case retrieval identifies prior judicial decisions relevant to a fact pattern, statutory retrieval locates applicable legislative provisions, and regulation retrieval seeks relevant administrative rules \cite{feng2024legalretrieval, statutoryretrieval}. 

Recent work adopts agentic system designs to support high-quality legal retrieval. First, structured query generation enable LLM agents to translate underspecified user inputs into legally grounded retrieval queries, with mechanisms such as candidate buffers rewriting queries and preserving intermediate results across turns to improve retrieval stability and success rates \citep{liu2022query}. Second, robustness is improved through multi-granular and logic-aware fusion of legal information, where documents are modeled at different levels such as facts, reasoning, and rulings, enabling LLM agent systems to tolerate paraphrase and noisy expressions while maintaining legal relevance \citep{meng2025multi}. Finally, multi-agent and neural-symbolic workflows are introduced to enhance verification and trustworthiness. Orchestrated agent systems explicitly check jurisdictional and temporal validity before producing answers \citep{wang2025mars}, while neural-symbolic approaches formalize statutes and use symbolic solvers to provide machine-checkable support for retrieved and applied legal rules \citep{chenTrustworthyLegalAI2025}.

\paragraph{Legal Theory Understanding \& Legal Reasoning.}

Legal theory defines the set of concepts and rules that underpin legal claims and their justification, including elements of offenses, duties, rights, and how these elements relate to one another \citep{yuan2024can, sadowski2025verifiable}. Legal reasoning is the process of applying legal theory to facts using deduction, analogy, and rule application. Recent work shows that standalone LLMs do not reliably grasp legal theories or their formal structure \citep{yuan2024can, jing2025maslegalbench}, and they also struggle to perform reliable legal reasoning without extra structure or checks \citep{jing2025maslegalbench, sadowski2025verifiable}.

To improve legal reasoning and theory understanding, recent work follows complementary agentic design strategies.  
First, multi-agent decomposition is used to split complex tasks into focused sub-tasks, enabling role specialization.  Systems such as MALR and ChatLaw show that a multi-agent design helps models focus on theory extraction, fact analysis, and synthesis separately \citep{yuan2024can, cui2023chatlaw}.  
Second, some systems decouple retrieval and reasoning into atomic primitives that a planner agent composes; the Deterministic Legal Agents work proposes a canonical primitive API for auditable, point-in-time retrieval and deterministic steps \citep{de2025deterministic}. 
Third, a reflective multi-agent framework is used for legal argument generation in which specialized agents iteratively analyze facts and refine arguments, improving factual grounding, ethical reliability, and robustness of legal reasoning compared to single-agent approaches \citep{zhang2025mitigating}.  

\subsection{Litigation \& Dispute Resolution}
Litigation and dispute resolution covers legal tasks that arise in adversarial proceedings, which involve analyzing case facts, applying relevant law, and engaging in strategic interactions between opposing parties and judges. Core litigation-related tasks include predicting judicial outcomes, drafting litigation documents such as motions and briefs, and simulating courtroom or mediation processes.

\paragraph{Judgment Prediction.}
On the task of legal judgment prediction, \citet{shui2023comprehensive} suggests that effective legal judgment prediction requires more sophisticated mechanisms beyond naive LLM–IR pipelines. Building on this limitation, \citet{jiangAgentsBench2025} propose a multi-agent LLM framework that explicitly models judicial deliberation through role-based agents, where individual agents independently reason over cases and iteratively reach a consensus through discussion, leading to consistent improvements in both predictive accuracy and decision quality. Further advancing agent-based legal reasoning, \citet{chen2025debate} introduce a Debate-Feedback framework that augments multi-agent debate with explicit reliability evaluation, mitigating the instability of uncontrolled debate and yielding more stable and effective legal judgment prediction.

\paragraph{Courtroom \& Dispute Resolution Simulation.}
Courtroom simulation must handle dynamic courtroom interactions, including multi-round adversarial debates, real-time responses, and evolving argument strategies, motivating agent-based frameworks with role-specialization and long-horizon interactions. Early works such as \textsc{AgentCourt} introduces adversarial lawyer agents to simulate courtroom debate and enable automated knowledge evolution through interaction \citep{chenAgentCourtSimulatingCourt2025a}. Subsequent work further integrate egal knowledge augmentation and structured court debate to improve judicial decision-making fidelity \cite{he-etal-2024-agentscourt}. Related systems explore courtroom reasoning from different perspectives: \textsc{MASER} simulates scalable synthetic legal scenario data by simulating interactions between roles such as client, lawyer, and supervisor \cite{shengbinyue2025multi}.

Beyond generic courtroom debate, several works focus on jurisdiction- or process-specific simulation. \textsc{Simcourt} introduces a multi-agent framework to simulate real and well-structured trial procedures of Chinese criminal courts, showing potentials to outperform human legal practitioners \citep{zhang2025chinese}. Broader institutional simulations further extend this line of work: \textsc{SAMVAD} models judicial deliberation dynamics in India \citep{devadiga2025samvad}, while \textsc{LegalSim} simulates legal systems to uncover procedurally valid yet harmful strategies, enabling stress-testing of legal processes \citep{badhe2025legalsim}. Besides court simulation, agentic simulation has also been applied to adjacent dispute-resolution settings. \textsc{AgentMediation}, simulates full five-stage civil dispute mediation with configurable roles, enabling experiments that reveal patterns like group polarization and surface consensus \citep{chen2025simulating}.

\subsection{Compliance, Governance, \& Regulation}

Compliance, governance and regulation aim at ensuring that individuals and organizations adhere to applicable laws, regulations, and governance standards.
These tasks are largely proactive and continuous, and often involve analyzing legislation and regulatory guidance, translating legal requirements into organizational policies, and ensuring accountability through audits and oversight.

\paragraph{Compliance.}
In financial crime prevention, agentic architectures assign bounded roles to autonomous agents, emphasizing compliance-by-design through traceability, explainability, and audit logging \citep{axelsen2025agentic}. For enterprise compliance, conversational agentic
assistant \textsc{CBA} routes user queries between a fast RAG-based mode and a full agentic mode, balancing latency and capability by dynamically selecting execution paths \citep{zhu2025compliance}. Embeddings-driven graph enable agentic systems such as \textsc{PACT}\cite{fang2025privacy} to link heterogeneous enterprise artifacts over metadata, ownership, and compliance context and reason over privacy-relevant connections at scale. Complementary work proposes specialized agentic systems that function as ethics counsel within legal workflows, providing accountable guidance and mitigating bias in professional decision-making \citep{o2024agentic}. 

Agentic approaches have also been explored in data-intensive compliance settings. For instance, a multi-agent system assists drug asset in due diligence, combining a web-browsing agent with an LLM-as-a-judge to suppress hallucinations and improve precision \citep{vinogradova2025llm}.
For data protection and privacy compliance, multi-agent systems decompose complex regulatory obligations into planning, execution, and verification roles to support end-to-end governance and data transfer validation \citep{nguyen2025multi, videsjorden2025positioning}.
At the regulatory level, multi-agent RAG frameworks construct structured representations of regulations to enable traceable, ontology-free compliance question answering \citep{agarwal2025ragulating}. Beyond enterprise settings, LLM agent moderation system has been applied to identifying non-compliant content in decentralized social platforms \cite{la2025safeguarding}, and to context-aware, jurisdiction-specific child safety moderation \cite{fillies2025scalable}.

Additionally, commercial AI agents are emerging to automate regulatory compliance workflows. \textsc{Regology} provides agents for regulatory research and change management~\citep{Regology}, while \textsc{V7Labs} offers regulatory cross-referencing and compliance verification~\citep{v7labs2025}. \textsc{Akira AI} automates regulatory monitoring, risk analysis, and policy adjustments~\citep{Akira}.

\paragraph{Risk Assessment \& Fraud detection.}
Recent research utilizes multi-agent frameworks to break down risk assessment and fraud detection tasks into specialized roles. For example, \citet{park2024enhancing} and \textsc{HCLA} \citep{na2025human} assign agents to data transformation, anomaly identification, and reporting. Other systems, such as \textsc{FinCon} and \textsc{MASCA}, integrate risk-control agents directly into financial decision-making and credit assessment pipelines, embedding risk modeling into the agent's workflow  \citep{yu2024fincon, jajoo2025masca}.
Beyond detection, agentic designs are used to enforce security and evaluate systemic hazards. A robotic security framework \citep{shah2025enforcing} combines LLM agents with blockchain and monitoring to safeguard online transactions. To ensure system safety, \textsc{M-SAEA} employs safety-aware auditor agents to probe multi-agent teams for coordination and deployment risks \citep{chen2025tasks}. \textsc{Co-Investigator AI} divides Anti-Money Laundering tasks into planning, detection, and validation stages, coordinate through shared memory and use an Agent-as-a-Judge loop to ensure regulatory accuracy with human oversight \citep{naik2025co}.

For insider threats and phishing, \textsc{Audit-LLM} \citep{song2024audit} and \textsc{GABM} \citep{ferraro2025generative} use Decomposer and Supervisor agents to analyze logs for insider threats. For cyber-attacks, \textsc{MultiPhishGuard} \citep{xue2025multiphishguard} and \textsc{PhishDebate} \citep{li2025phishdebate, nguyen2025debate} utilize debating and judge agents to aggregate evidence against deceptive URLs and emails. Additionally, such agentic internal control extends to other fields:  \citet{gu5561180llm} detects greenwashing, \textsc{WandaPlan} \citep{yao2025your} identifies misinformation, \textsc{SafeScientist} \citep{zhu2025safescientist} monitors ethical risks in science, \textsc{Agent4FaceForgery} \citep{lai2025agent4faceforgery} uses memory-equipped agents to detect forged facial data, and \citet{islayem2025using} introduces a blockchain–LLM agentic framework for detecting health insurance fraud through smart-contract enforcement and retrieval-grounded analysis.

\paragraph{Governance, Regulation \& Legislation.}
In the governance domain, Agentic legal assistants integrate retrieval-augmented generation with LLM agents to support interactive querying and reasoning over EU GDPR legislation and case law \citep{mamalis2024large}. Further, multi-agent simulations are used to model governance dynamics: \textsc{Law in Silico} combines individual and institutional agents to simulate lawmaking, adjudication, and enforcement \citep{wang2025law}. Other frameworks simulate cooperative governance in shared-resource settings, showing how planning, communication, and moral reasoning shape collective outcomes \citep{piatti2024cooperate, zhou2025llm}.

For legislative analysis and simulation, prior work models U.S. Senate committee discussions using LLM-driven agents to study debate, reflection, and bipartisan decision-making under controlled conditions \citep{baker2024simulating}. \textsc{LegalSim} introduces a multi-agent legal simulation to uncover procedurally valid yet harmful strategies, enabling stress-testing of legislation \citep{badhe2025legalsim}. Other systems simulate legislative behavior to predict roll-call votes using role-based LLM agents with interpretable reasoning \citep{li2025political}.

\subsection{Advisory Consultation \& Transactions}

Advisory consultation focuses on responding to consultation-style inquiries, requiring the translation of legal rules into practical guidance. 
Transactional practice, by contrast, centers on the analysis and negotiation of exchanges of rights, obligations, and assets among multiple parties. Core transactional tasks include contract review and drafting and negotiation of deal terms. Transactional work often extends to specialized domains such as patent registration, real estate, bankruptcy, and merger and acquisition (M\&A), where domain-specific rule and procedural constraints are critical.

\paragraph{Query-Based Advisory Consultation.}
Recent research adopts agentic, multi-agent LLM systems to support reliable responses to user-posed legal queries and consultation-style questions.
\textsc{ChatLaw}~\cite{cui2023chatlaw} treats legal consultation as a series of structured question-answering steps, using role-specialized agents and a knowledge-graph–enhanced mixture-of-experts to reduce hallucinations in practical legal Q\&A.
Similarly, \textsc{LawLuo}~\cite{sun2024lawluo} uses role-based agents to support multi-round Chinese legal consultations, iteratively clarifying and answering user queries to produce structured consultation reports.
To improve trustworthiness in high-stakes question answering, Xi et al.~\cite{xi2025hybridrag} propose an agentic hybrid RAG framework that dynamically routes legal queries between retrieval-based answering and multi-model generation, using a selector agent to choose the most reliable response. 
 
\textsc{L-MARS}~\cite{wang2025mars} combines agentic search with multi-agent reasoning to answer user queries with improved factuality and reduced uncertainty.
Finally, \textsc{LegalGPT}~\cite{shi2024legalgpt} formalizes legal chain-of-thought reasoning within a multi-agent framework for query answering.

\paragraph{Contracts, Negotiation, and Transactions.}
For contract review and drafting, systems such as \textsc{PAKTON}~\cite{raptopoulos2025pakton} and \textsc{LAW}~\cite{watson-etal-2025-law} frame contract question answering and analysis as role-structured multi-agent pipelines, enabling traceable reasoning over long agreements.
Complementary retrieval-augmented multi-agent systems automate the drafting of transactional documents by combining drafting, validation, and compliance agents~\citep{suravarjhula2025SOW}. \textsc{Rawlsian Agents} applies agentic reasoning grounded in fairness principles to evaluate and draft bilateral agreements \citep{ferrorawlsian}. In addition to academic advances, industrial practice has increasingly adopted AI agents for contract review and drafting. Top international law firms are collaborating with legal AI companies to incorporate agentic AI for complex legal tasks such as reviewing loan documents for leveraged finance transactions \cite{AOHarvey, Pearson}. Similarly, \textsc{Spellbook AI Associate} provides agentic services covering streamlined agreement revision for commercial lawyers \cite{Spellbook}. Other non-legal focused AI agent startups also increasingly expand their services and products into contract review and analysis domains by providing customizable legal agents to assist in contract drafting \citep{Harvey},  extracting key contract clauses to streamline contract reviews \cite{odinai2025}, and autonomous contract revision and drafting \cite{zbrain, Flank, Aline, Legora, Spellbook}. 

In the context of negotiation, agentic systems investigate autonomous bargaining and deal-making behaviors. Work such as \textsc{NegotiationGym}~\cite{mangla2025negotiationgym} simulates iterative negotiation dynamics among self-optimizing agents. Other studies use large-scale simulations to study strategic dynamics and risks in agent-to-agent negotiations and consumer transactions~\cite{vaccaro2025advancing, zhu2025automated}. In addition to contract and business negotiation, some recent works focus on specialized transactional domains including agentic systems for automated patent analysis, filing, and document drafting~\cite{wang2024patentagent,wang2024autopatent,shea2025autospec,sakhinana2024towards}, real estate transactions and housing services~\cite{haurum2024real}, corporate bankruptcy analysis~\citep{maturo2025arcadia}, tax preparation~\citep{gogani2025tax}, and M\&A evaluation~\citep{mirzayev2025aMA}.

\subsection{Workflow Automation}

Non-substantive tasks are also essential for the legal industry, which focus on supporting internal operations rather than legal analysis itself.
Legal practitioners and vendors use AI agents to automate repetitive and high-volume steps around intake and internal measures. For instance, intelligent intake agents collect basic case information and schedule consultations by interacting with calendaring systems \citep{voiceflow2025}. Email-management agents support inbox triage, summarization, and draft generation, helping lawyers manage high volumes of correspondence more efficiently \citep{jace2025, relevanceai}. Legal AI agents also assist in monitoring system states, detecting changes, and triggering actions across tools without requiring constant user prompts \citep{mehrjardi2025}.  

Workflow automation agents are also applied to internal coordination tasks such as document routing and version control. \textsc{AgentiveAIQ} enable firms to build branded agents that combine retrieval-augmented generation with knowledge graphs to autonomously perform client triage and internal insights delivery \citep{agentiveaiq2025}. \textsc{AIQLabs} provides agentic services on operational services including invoice automation, knowledge management, and recruiting automation \citep{aiqlabs2025}. \textsc{Odin AI} offers a business AI agent focusing on meeting notes summarization and task automation \citep{odinai2025}. \textsc{LawBot AI} provides legal-specific voice assistant agents for automatically handling client inquiries, scheduling consultation, and delivering case information \citep{lawbot2025}.

\section{Discussions of Legal LLM Agent Evaluation}
Evaluating LLM agents for legal tasks requires criteria that go beyond standard language modeling metrics, and has shifted from single-turn, standalone LLM assessments to multi-step agent-based evaluations. 
Despite this progress, significant challenges remain, including designing dedicated metrics for nuanced legal concepts.
This section reviews recent work on common evaluation dimensions, benchmarks, and empirical results, and summarizes current progress of legal agent systems. Table~\ref{tab:legal-benchmarks-rotated} summarizes the commonly used legal benchmarks and their key properties. 

\subsection{Common Evaluation Criteria for the Legal Domain}
Legal validity assesses whether an LLM's output is legally sound and defensible within the applicable legal framework and is commonly evaluated through three core dimensions: substantive correctness, legal reasoning correctness, and ethics.

\paragraph{Substantive Correctness.}
Substantive correctness measures whether an LLM produces a legally correct conclusion under the applicable law given the facts, independent of its reasoning process. Existing work operationalizes this notion mainly through two paradigms: (i) task-based paradigms that assesses whether LLMs or agents successfully complete legal tasks with varying complexity using task completion or final-answer accuracy as the primary metric \cite{fei-etal-2024-lawbench, liLexEvalComprehensiveChinese, li2025legalagentbench, jiaReadyJuristOne2025}; and (ii) legal Q\&A settings, in which models' answers to legal questions are evaluated against expert-authored reference answers, using fine-grained relational judgments, such as equivalence or contradiction, to capture outcome-level correctness \citep{bhambhoriaEvaluatingAILaw2024}.

\paragraph{Legal Reasoning Correctness.}
Existing work evaluates the reasoning capabilities of legal agents by validating intermediate steps and structured argumentation, including keyword-matching rates in function-calling outputs \cite{li2025legalagentbench}, and the Issue–Rule–Application–Conclusion (IRAC)-based decomposition of legal reasoning into structured and verifiable stages \cite{guhaLegalBenchCollaborativelyBuilt2023}. Other studies assess syllogistic reasoning by examining the logical coherence and alignment of LLM-generated responses, premises, and legal principles, jointly evaluating both reasoning correctness and citation quality \cite{zhangCitaLawEnhancingLLM2025}. Similarly, \citet{daiLAiWChineseLegal2025} categorize legal reasoning capabilities into three progressively complex levels: fundamental information retrieval, legal principle inference, and advanced legal applications.

\paragraph{Ethics.}
Existing work also emphasizes the need to evaluate whether LLMs behave in ethically appropriate ways in legal contexts, focusing on whether model outputs conform to professional norms, legal obligations, and societal expectations, particularly in settings involving legal advice or decision support, where risks include misleading guidance, unauthorized practice of law, and biased or discriminatory behavior.

Several studies argue that such evaluation requires domain-specific metrics rather than generic AI ethics principles. \citet{wrightAILawAssessing2020} notes that existing ethical guidelines are often high-level and insufficient for assessing concrete system behavior in legal practice, and calls for context-sensitive, measurable criteria that reflect how AI systems are deployed and used. \citet{zhangEvaluationEthicsLLMs2024} propose an operational framework for ethical evaluation in legal LLMs, introducing dimensions such as legal instruction following, legal knowledge consistency, and robustness to misleading or adversarial prompts. 

\subsection{Benchmarks Overview}
Table \ref{tab:legal-benchmarks-rotated} summarizes a diverse set of benchmarks for evaluating LLM-based agents on legal tasks. These benchmarks span procedural accuracy, legal reasoning, judgment quality, ethics, and fairness across simulated and real-world settings. Evaluation methods include rule-based metrics, statistical models, human annotation, and LLM-as-judge frameworks. Task formats range from single-turn classification to multi-agent simulations and adversarial court scenarios: \textsc{CourtReasoner} examines whether LLM agents can reason like judges through structured deliberation \citep{han-etal-2025-courtreasoner}; to evaluate agent-based court simulation systems, \textsc{SimuCourt} provides a benchmark comprising 420 real-world Chinese judgment documents across criminal, civil, and administrative cases in first- and second-instance settings \citep{he-etal-2024-agentscourt}; \textsc{MASLegalBench} provide strict, deductive testbeds to compare multi-agent designs and measure progress on formal legal reasoning tasks \citep{jing2025maslegalbench}.

\subsection{Fidelity of Evaluation Metrics}
Many existing evaluation benchmarks rely heavily on human-authored reference data, which is costly and difficult to scale. To reduce annotation overhead, recent work has proposed \emph{LLM-as-a-judge} approaches that automate evaluation by using LLMs to assess generated outputs. However, this paradigm raises concerns about \emph{fidelity}, defined as the degree to which an evaluation metric aligns with how legal experts assess quality. Low-fidelity evaluation can undermine both the reliability and the practical effectiveness of benchmarking results. 

Several studies have proposed methods to improve fidelity by better approximating expert legal judgment. For example, \citet{enguehardLeMAJLegalLLMasaJudge2025} decompose long-form legal answers into fine-grained \emph{legal data points}, enabling evaluation at a level that more closely reflects how lawyers assess correctness and omissions. Other work incorporates external legal context into the evaluation process: retrieval-based evaluation methods for legal QA, where model outputs are judged against retrieved supporting documents, improving correlation with lawyer assessments relative to purely model-based evaluators, and reducing reliance on surface-level similarity \citep{ryuRetrievalbasedEvaluationLLMs2023}. In addition, \citet{rajuConstructingDomainSpecificEvaluation2024} improve evaluation fidelity by constructing domain-specific benchmarks and data pipelines that better align model assessment with human preferences. Beyond that,\citet{cheongAIAmNot2024} focus more on the practical side, and expert-centered studies show that appropriate LLM systems may prioritize helping users ask the right questions and avoid unauthorized practice of law, implying that efficacy must be assessed relative to realistic legal-use objectives rather than only answer correctness.

\subsection{Challenges in Evaluation}
We identify several key challenges in the current evaluation of legal LLMs and agent-based systems:
\begin{itemize}[leftmargin=*]\setlength\itemsep{-0.3em}
    \item \emph{Metric-related limitations.} Existing evaluations largely lack coverage of law-specific, substantively important metrics, such as procedural compliance and bias or fairness assessment. Only a limited number of studies explicitly address these dimensions. For example, \cite{jiaReadyJuristOne2025} evaluates procedural correctness, while \cite{xueJustEvaToolkitEvaluate2025} examines judicial fairness and bias. Given the centrality of such properties to legal reasoning and decision-making, more systematic investigation of law-specific evaluation metrics is urgently needed.

    \item \emph{Distributional bias across legal systems.} Current benchmarks and empirical studies exhibit limited diversity in legal systems and traditions. We observe that most existing datasets and evaluations are grounded in the Chinese legal system, with relatively few benchmarks derived from U.S., European, or other legal jurisdictions. Because legal rules, procedures, and normative standards vary substantially across jurisdictions, conclusions drawn from a single legal system may not generalize and may introduce systemic bias.

    \item \emph{Evaluation fidelity in agent-based settings.} Many agent-based evaluations rely on LLM-driven role-play to simulate legal actors such as lawyers or judges (e.g., \cite{chenAgentCourtSimulatingCourt2025a, jiaReadyJuristOne2025}). However, the realism, stability, and reliability of these simulated agents remain insufficiently validated. It is  unclear to what extent their behavior reflects real-world legal practice, how sensitive they are to prompt variations, or how their limitations may confound evaluation outcomes.

    \item \emph{Limited evaluation of multi-agent systems.} Although recent work has begun to explore multi-agent legal systems (e.g., \cite{jiangAgentsBench2025, cui2023chatlaw, chenTrustworthyLegalAI2025}), dedicated evaluation frameworks for multi-agent settings are still lacking. In particular, reliability-oriented dimensions, such as workflow coherence, agent coordination, and inter-agent consistency, are rarely assessed in a principled manner.

    \item \emph{Fairness in prompt tuning and agentic framework adaptation.} Many evaluations do not adequately account for model-specific prompt tuning or agentic framework adaptation, which can lead to suboptimal performance and undermine the validity of cross-model comparisons. While some studies explore zero-shot and few-shot prompting strategies for LLMs (e.g., \cite{fei-etal-2024-lawbench, liLexEvalComprehensiveChinese}) or vary agent architectures to obtain a more comprehensive view \cite{li2025legalagentbench}, there is no widely accepted standard for ensuring fair and comparable evaluation across models. Given the sensitivity of LLM behavior to prompting and system design, standardized protocols are needed to support meaningful comparative conclusions.
\end{itemize}

\section{Current Progress, Challenges and Future Directions}

\subsection{Progress and Remaining Challenges}

On the technical side, existing benchmarks evaluate LLM agents from procedural accuracy, legal reasoning, judgment quality, ethics, and fairness across simulated and real-world settings \cite{he-etal-2024-agentscourt, han-etal-2025-courtreasoner, jing2025maslegalbench}.  These evaluation results reveal strong performance for information retrieval with clear targets and template-constrained drafting.

Despite the achievements, recent evaluation results still reveal persistent weaknesses in LLM agents across critical legal tasks. First, sensitivity to facts and perturbations is a notable limitation. When key facts are removed or altered in judicial reasoning benchmarks, agents often fail to adjust reasoning appropriately \cite{han-etal-2025-courtreasoner}. This brittleness under small factual changes suggests that agents rely excessively on surface patterns rather than robust doctrinal understanding \cite{han-etal-2025-courtreasoner}.
Second, Procedural and dynamic task failure is another core challenge. Agents fall short of realistic professional competency, especially in high-complexity procedural settings and when procedural correctness is paramount \citep{jiaReadyJuristOne2025}. 
Further, evaluating legal LLMs and agentic systems remains challenging due to the insufficient law-specific metrics, limited jurisdictional diversity, and insufficient validation of agent-based simulations. Existing benchmarks often under-evaluate procedural correctness, fairness, and multi-agent coordination \citep{jiaReadyJuristOne2025, xueJustEvaToolkitEvaluate2025}, while reliance on prompt-driven role-play raises concerns about stability and reproducibility \citep{chenAgentCourtSimulatingCourt2025a, jiaReadyJuristOne2025}. 

Beyond technical limitations, deploying LLM agents in legal contexts raises substantial legal and ethical concerns. Issues such as alignment with professional responsibility rules remain underexplored, especially when agents act autonomously or interact directly with clients \citep{o2024agentic, agarwal2025ragulating}. These concerns underscore the need for careful governance, human oversight, and compliance-aware agent design before large-scale deployment.

\subsection{Future Directions}

Future research on LLM agents for legal tasks should prioritize designs that better align with legal practice requirements. A key direction is integrating human expertise in the loop, where legal professionals supervise and review agent outputs before deployment or downstream use. Human oversight has been shown to improve factual accuracy, procedural compliance, and alignment with professional standards, particularly for high-stakes legal decisions and multilingual or jurisdiction-sensitive settings \citep{meng2025building, o2024agentic}.
Another critical direction is the development of new benchmarks and evaluation protocols tailored to agentic legal systems. Existing closed-form benchmarks are insufficient for capturing long-horizon execution, interaction quality, and procedural correctness. Future evaluations should incorporate interactive environments, reasoning trace validation, and mixed automatic–human assessment to better reflect real-world legal workflows \citep{jing2025maslegalbench, jiaReadyJuristOne2025, li2025legalagentbench}. Finally, robust governance models and verification protocols to manage agentic AI in the legal field are critical. Policy frameworks should establish accountability mechanisms, such as requiring human lawyers' sign-off when agentic AI is used in real-life legal tasks.  \cite{kraprayoon2025ai}

\section{Conclusions}
In this survey, we provided a comprehensive examination of LLM agents as a solution for complex legal tasks.
We systematically analyzed the technical transition toward agentic architectures and presented a structured taxonomy of applications ranging from litigation assistance to workflow automation. While these agents effectively mitigate persistent issues like hallucination and outdated information through modular execution and tool usage, several challenges regarding long-horizon reliability, procedural correctness, and citation fidelity remain. Lastly, we outline promising directions for future research.


\section*{Limitations}
This survey has several limitations. First, our survey focuses primarily on LLM-based agents for legal tasks, and our coverage skews toward English-language publications and systems. Although we tried to cover varying jurisdictions, including common law jurisdictions (e.g. the US and UK), India, and Chinese legal systems. However, the scope may potentially under represents non-English paper or products in other regions.
Second, our coverage of commercial legal AI agents relies primarily on publicly available information such as press releases, product documentation, and vendor websites. We lack access to proprietary system architectures, internal evaluations, or user adoption data. Consequently, our characterization of commercial systems may be incomplete, as we cannot independently verify vendor claims about system capabilities or performance.
Third, given the complexity of legal work, distinctions between task categories are not always clear-cut, and real-world practice often encompasses multiple task types simultaneously. Although our categorization of legal domains and tasks is grounded in established legal practice, the boundaries between categories such as “litigation support” and “legal consultation,” or “compliance” and “fraud detection,” are inherently fluid, and certain systems may reasonably be classified under multiple categories.

\bibliographystyle{plainnat}
\bibliography{main}

\clearpage

\appendix
\onecolumn 
\section{Legal Agentic Systems}

\small
\renewcommand{\arraystretch}{1.15}
\begin{longtable}{
    p{2.25cm}    
    p{2.7cm}      
    p{3cm}      
    p{3.5cm}    
    p{2.5cm}   
}
\caption{Comprehensive overview of representative LLM agents for legal tasks (literature and academic report).}
\label{tab:legal_llm_agents} \\

\toprule
\textbf{Domain} &
\textbf{Agentic System} &
\textbf{Author, Year} &
\textbf{Agentic Design Pattern} &
\textbf{Task} \\
\midrule
\endfirsthead

\multicolumn{5}{c}{\tablename\ \thetable{} -- \textit{Continued from previous page}} \\[2pt]
\toprule
\textbf{Domain} &
\textbf{Agentic System} &
\textbf{Author, Year} &
\textbf{Agentic Design Pattern} &
\textbf{Task} \\
\midrule
\endhead

\midrule
\multicolumn{5}{r}{\textit{Continued on next page}} \\
\endfoot

\bottomrule
\endlastfoot

\multirow{4}{*}{\parbox{2.8cm}{Legal Retrieval}} 
  & \textsc{L-MARS} & \citet{wang2025mars} & Reasoning–search–verification loop &  Multiple choice legal question \\
  & \textsc{L4M} & \citet{chenTrustworthyLegalAI2025} & Neural-symbolic workflow & Chinese legal case retrieval \\
\midrule

\multirow{11}{*}{\parbox{2.8cm}{Legal Reasoning}} 
  & \textsc{L-MARS} & \citet{wang2025mars} & Reasoning–search–verification loop &  Multiple choice legal question \\
  & \textsc{MLTDB} & \citet{meng2025building} & Human-in-the-loop on multi-agent & Multilingual legal terminology extraction \\
  & \textsc{Reflective Multi-Agent} & \citet{zhang2025mitigating} & Reflective multi-agent & Legal Argument Generation \\
  & \textsc{Deterministic Legal Agents} & \citet{de2025deterministic} & Canonical primitive API, point-in-time retrieval & Legal Q\&A\\
  & \textsc{MALR} & \citet{yuan2024can} & Non-parametric learning multi-agent & Confusing Charge Prediction Task \\
\midrule

\multirow{3}{*}{\parbox{2.8cm}{Judgment Prediction}} 
  & \textsc{Debate-Feedback} & \citet{chen2025debate} & Multi-agent debate \& verification & Trial Prediction \& Article Prediction \\
  & \textsc{AgentsBench} & \citet{jiangAgentsBench2025} & Multi-agent simulation of judicial bench's discussion & Criminal Prison-Term Prediction\\ 
\midrule

\multirow{7}{*}{\parbox{2.8cm}{Litigation \&\\Dispute Resolution}} 
  & \textsc{AgentCourt} & \citet{chenAgentCourtSimulatingCourt2025a} & Adversarial evolutionary multi-agent simulation & Civil Court Debate \& Legal Reasoning \\
  & \textsc{AgentsCourt} & \citet{he-etal-2024-agentscourt} & Multi-agent simulation of trial procedures & Judicial Decision-Making Process \\
  & \textsc{SimCourt} & \citet{zhang2025chinese} & Multi-agent simulation of trial procedures & Criminal Judgement Prediction \& Trial Process\\ 
  & \textsc{SAMVAD} & \citet{devadiga2025samvad} & Multi-agent simulation of trial procedures & Criminal Judgement Prediction \& Trial Process \\
  & \textsc{LegalSim} & \citet{badhe2025legalsim} & Multi-agent simulation of adversarial legal proceedings & Litigation procedural strategy discovery \\
  & \textsc{AgentMediation} & \citet{chen2025simulating} & Multi-agent simulation of dispute mediation process & Civil Dispute Mediation\\
  & \textsc{MASER} & \citet{shengbinyue2025multi} & Multi-agent simulation of legal service interactions & Legal consultation \& Complaint drafting \\
\midrule

\multirow{3}{*}{\parbox{2.8cm}{Compliance}} 
  & \textsc{CBA} & \citet{zhu2025compliance} & Combination of RAG and agentic mode & Legal Q\&A\\
  & \textsc{PACT} & \citet{fang2025privacy} & Embeddings-driven graph & Enterprise search \\
\midrule

\multirow{12}{*}{\parbox{2.8cm}{Fraud Detection}} 
  & \textsc{HCLA} & \citet{na2025human} & Multi-role conversational agents & Anomalous transaction detection \\
  & \textsc{M-SAEA} & \citet{chen2025tasks} & Multi-agent system with auditor role & Financial risk evaluation \\
  & \textsc{Co-Investigator AI} & \citet{naik2025co} & Multi-role agents and Agent-as-a-Judge loop & Anti-Money Laundering compliance \\
  & \textsc{FinCon} & \citet{yu2024fincon} & Multi-agent system with risk control module & Financial decision making \\
  & \textsc{MASCA} & \citet{jajoo2025masca} & Multi-agent system with risk control module & Credit assessment \\
  & \textsc{Audit-LLM} & \citet{song2024audit} & Multi-agent system for detection simulation & Insider threat detection \\
  & \textsc{GABM} & \citet{ferraro2025generative} & Hierarchical pipeline with specialists and supervisor & Insider threat detection \\
  & \textsc{MultiPhishGuard} & \citet{xue2025multiphishguard} & Multi-role agents and adversarial simulation & Phishing detection \\
  & \textsc{PhishDebate} & \citet{li2025phishdebate} & Multi-role debating agents & Phishing detection \\
  & \textsc{WandaPlan} & \citet{yao2025your} & Multi-role agentic pipeline & Misinformation and fraud detection \\
  & \textsc{SafeScientist} & \citet{zhu2025safescientist} & Research simulation with defense and attack modules & Scientific discovery safety\\
  & \textsc{Agent4FaceForgery} & \citet{lai2025agent4faceforgery} & Behavior simulation and adaptive rejection sampling   & Face forgery detection \\
\midrule

\multirow{5}{*}{\parbox{2.8cm}{Regulation}} 
  & \textsc{Law in Silico} & \citet{wang2025law} & Multi-agent law system simulation & Legal theory study \\
  & \textsc{GovSim} & \citet{piatti2024cooperate} & Cooperation simulation and moral reasoning & Social sustainability study \\
  & \textsc{LegalSim} & \citet{badhe2025legalsim} & Multi-agent pipeline for legal system simulation & Legislation improvement \\
  & \textsc{Political Actor Agent} & \citet{li2025political} & Legislative behavior simulation with role-playing agents & Roll-call vote prediction \\
  & \textsc{Rawlsian Agents} & \citet{ferrorawlsian} & Bilateral agent negotiation under law theory & Contract negotiation \\
\midrule

\multirow{4}{*}{\parbox{2.8cm}{Legal Consultation}} 
  & \textsc{ChatLaw} & \citet{cui2023chatlaw} & Role-specialized workflow (MoE) & Legal Q\&A \\
  & \textsc{LawLuo} & \citet{sun2024lawluo} & Multi-role conversational agents & Legal Q\&A \\
  & \textsc{LegalGPT} & \citet{shi2024legalgpt} & Chain-of-thought orchestration with role-based agents & Legal Q\&A \\
\midrule

\multirow{4}{*}{\parbox{2.8cm}{Contract Analysis}} 
  & \textsc{PAKTON} & \citet{raptopoulos2025pakton} & Document-centric agent pipeline & Contract / legal agreements \\
  & \textsc{LAW} & \citet{watson-etal-2025-law} & Legal agentic workflow for contracts  & Custody and fund services contract analysis \\
\midrule

\multirow{6}{*}{\parbox{2.8cm}{Transaction}} 
  & \textsc{NegotiationGym} & \citet{mangla2025negotiationgym} & Multi-agent negotiation simulation & Business negotiation \\
  & \textsc{ARCADIA} & \citet{maturo2025arcadia} & Agentic causal reasoning pipeline & Corporate bankruptcy analysis \\
  & \textsc{PatentAgent} & \citet{wang2024patentagent} & Role-specialized patent analysis agents & Patent application \& filing \\
  & \textsc{AutoPatent} & \citet{wang2024autopatent} & Multi-agent patent drafting pipeline & Patent application \& filing \\
  & \textsc{AutoSpec} & \citet{shea2025autospec} & Agentic specification generation workflow & Patent application \& filing \\
  & \textsc{Synedrion} & \citet{gogani2025tax} & Multi-agent deliberation and synthesis & Collaborative decision-making and agreement synthesis \\

\end{longtable}
\normalsize

\vspace{2mm}

\clearpage

\small
\renewcommand{\arraystretch}{1.15}
\begin{longtable}{
    p{2cm}    
    p{2.5cm}      
    p{2.5cm}    
    p{6.5cm}    
}
\caption{Commercial AI agent products for legal tasks (industrial products).}
\label{tab:commercial_legal_ai_agents} \\

\toprule
\textbf{Task Category} &
\textbf{Product / Company} &
\textbf{Website} &
\textbf{Use Cases} \\
\midrule
\endfirsthead

\multicolumn{4}{c}{\tablename\ \thetable{} -- \textit{Continued from previous page}} \\[2pt]
\toprule
\textbf{Task Category} &
\textbf{Product / Company} &
\textbf{Website} &
\textbf{Use Cases} \\
\midrule
\endhead

\midrule
\multicolumn{4}{r}{\textit{Continued on next page}} \\
\endfoot

\bottomrule
\endlastfoot

\multirow{5}{*}{\parbox{2.8cm}{Comprehensive \\ Legal Tasks }} 
  & \textsc{Harvey Agent} & \citet{Harvey} & Multi-model agents for legal research, contract review, due diligence, litigation support \\
  \addlinespace
  
  & \textsc{CoCounsel Legal} & \citet{Cocounsel} & Legal research by Deep Research with agentic planning, drafting with guided workflows, bulk document review \\
  \addlinespace
  
  & \textsc{LexisNexis Protégé} & \citet{Protege} & Legal research, drafting, document analysis through multiple agents \\
  \addlinespace
  
  & \textsc{Vincent} & \citet{Vincent} & Legal research, litigation, transactions through Agentic Workflow Engine with 20+ pre-built workflows, cross-jurisdictional reasoning  \\
  \addlinespace
  
  & \textsc{Luminance} & \citet{Luminance} & Autonomous contract review, compliance analysis, negotiation \\
  \addlinespace

\midrule

\multirow{2}{*}{\parbox{2.8cm}{Legal Search}} 
  & \textsc{DeepJudge} & \citet{Deepjudge} & Legal reasoning and multi-step legal search or information retrieval  \\
  \addlinespace
\midrule

\multirow{2}{*}{\parbox{2.8cm}{Litigation}} 
  & \textsc{Clearbreif} & \citet{Clearbrief} & Litigation focused tasks; offering Microsoft Add-ons for legal document processing  \\
  \addlinespace
\midrule

\multirow{3}{*}{\parbox{2.8cm}{Compliance \& \\ Regulatory}} 
  & \textsc{Regology} & \citet{Regology} & AI agents for research, regulatory change, and compliance workflows  \\
  \addlinespace
  & \textsc{V7labs} & \citet{V7labs} & Regulatory cross-referencing and compliance verification \\
  \addlinespace
  & \textsc{Akira} & \citet{Akira} & Automates regulatory monitoring, analyzes risks, and ensures timely policy adjustments \\
\midrule

\multirow{8}{*}{\parbox{2.8cm}{Contract \& \\Transaction}} 
  
  & \textsc{Flank Agents} & \citet{Flank}  & Autonomous enterprise legal system covering reviewing contracts, answering compliance questions, and drafting legal documents \\
  \addlinespace
  & \textsc{Speelbook AI Associate}  & \citet{Spellbook} & Streamlined legal agreement/contract revision for commercial transactions \\
  \addlinespace
  & \textsc {Aline} & \citet{Aline} & Contract lifecycle and AI-assisted contract automation for in-house legal teams  \\
  \addlinespace
  & \textsc {Pearson} & \citet{Pearson} & Automating corporate transactions, including M\&A due diligence and financing\\
  \addlinespace
  & \textsc {Legora} & \citet{Legora} & Customizable agentic workflows for lawyers covering varying perspectives of commercial transactions \\
  \addlinespace
  & \textsc{LeAh} & \citet{Leah} & Contract lifecycle management, legal workflows, procurement through agentic AI platform \\
  \addlinespace
  & \textsc{Juro} & \citet{Juro} & Contract lifecycle management, contract drafting, negotiation through workflow-embedded playbook redlining, risk surfacing \\
  \addlinespace
  & \textsc{Lito} & \citet{Lito} & Contract review, due diligence, and contract summaries \\
\midrule

\multirow{4}{*}{\parbox{2.8cm}{Supporting\\Legal Tasks}} 
  & \textsc{AgentiveAIQ} & \citet{agentiveaiq2025}  & Client triage \& internal insights delivery \\
  \addlinespace
  & \textsc{AIQLabs} & \citet{aiqlabs2025}  & Invoice automation, knowledge mgmt, recruiting \\
  \addlinespace
  & \textsc{Odin AI} & \citet{odinai2025}  & Meeting notes summarization \& task automation \\
  \addlinespace
  & \textsc{LawBot AI} & \citet{lawbot2025} & Client inquiries, scheduling, case info delivery \\
  \addlinespace
  & \textsc{Caseflood.ai} & \citet{Caseflood.ai} & Administrative tasks including phone intake and client follow-up. \\

\end{longtable}
\normalsize

\vspace{2mm}
\twocolumn


\begin{sidewaystable}[p]
\scriptsize
\setlength{\tabcolsep}{3pt}
\renewcommand{\arraystretch}{0.92}

\captionsetup{skip=6pt}
\caption{Legal Benchmarks Summary}
\label{tab:legal-benchmarks-rotated}

\makebox[\textwidth][c]{%
\begin{minipage}{\textheight}
\centering
\begin{tabularx}{\linewidth}{
  L{2.4cm}
  L{2.2cm}
  L{2.6cm}
  Y
  Y
  L{2.6cm}
  L{2.6cm}
  L{2.4cm}
  L{1.3cm}
}
\toprule
\textbf{Benchmark} & \textbf{Paper} & \textbf{Construct} &
\textbf{Measurable} & \textbf{Metrics} &
\textbf{Size} & \textbf{Simulation} &
\textbf{Judge} & \textbf{Type} \\
\midrule
J1-EVAL / J1-ENVS &
\citet{jiaReadyJuristOne2025}&
Task performance and procedural compliance &
Task success; procedural correctness &
BIN; NBIN; PFS; JUD &
6 environments, 3 levels, 508 scenarios &
Multi-agent role-playing sandbox, multi-turn &
LLM-as-Judge (GPT-4o) + Rule-based &
Agent \\
\addlinespace
LegalAgentBench &
\citet{li2025legalagentbench} &
Practical legal agent capability of tooling and interaction &
Interaction success \& Process rate &
Success rate; BERT-Score &
300 tasks &
Single-turn, multi-step tool-augmented agent &
Rule-based (Keyword/Program) &
Agent \\
\addlinespace
CourtReasoner & \citet{han-etal-2025-courtreasoner} & LLMs’ legal reasoning quality based on real court opinion documents & Citation relevance, constraint extraction accuracy, argument validity evaluation & Human evaluation scores & 292 expert-annotated meta-evalaution examples & Single-turn & Human annotators & Agent \\
\addlinespace

SimuCourt & \citet{he-etal-2024-agentscourt} & Judicial decision-making & Legal article generation and judgment accuracy & Precision, recall and F1 scores & 420 real-world judgment documents & Multi-agent simulated court debate embedded in decision-making process & LLM-as-Judge + human evaluation & Agent\\
\addlinespace

MASLegalBench & \citet{jing2025maslegalbench} & Deductive legal reasoning with multi-agent systems tailored to GDPR and rich real-world legal contexts & Performance on legal sub-tasks & Human evaluation scores & ~950 legal questions built from expert-authored court case contexts & Multi-agent deduction tasks with role specialization and collaborative reasoning configurations & Human evaluation & Agent \\
\addlinespace

CourtBench &
\citet{chenAgentCourtSimulatingCourt2025a} &
Interactive Reasoning Capability &
Court performance over civil cases &
Case winning rate; Prof score &
1,000 civil cases &
Multi-turn, Adversarial Evolution &
LLM-as-judge &
Agent \\
\addlinespace
Trident-Bench &
\citet{huiTRIDENTBenchmarkingLLM2025} &
Domain Safety \& Ethics &
Adherence to Professional Codes &
Violation/Refusal Rate (\%) &
887 prompts &
Single-turn &
Rule-based &
LLM \\
\addlinespace
LAiW &
\citet{daiLAiWChineseLegal2025} &
Legal Syllogism &
Logic (FIR, LPI, ALP) &
Accuracy, F1, Macro-F1 &
11,000 tasks &
Single-turn &
Rule-based &
LLM \\
\addlinespace
JudiFair &
\citet{xueJustEvaToolkitEvaluate2025}&
Judicial Fairness \& Bias &
Extra-legal factor influence &
Inconsistency; Bias; Imbalanced Acc &
65 extra-legal labels &
Single-turn &
Rule-based (Statistical) &
LLM \\
\addlinespace
LegalEval-Q &
\citet{yunhanLegalEvalQNewBenchmark2025}&
Linguistic Quality &
Clarity, coherence, terminology &
AdjScore; 0--100\% &
946 annotated queries &
Single-turn &
Logistic Regression &
LLM \\
\addlinespace
LexEval &
\citet{li2025legalagentbench} &
Cognitive Ability &
LexAbility Taxonomy &
Accuracy, ROUGE-L &
14,150 questions &
Single-turn &
Rule-based &
LLM \\
\addlinespace
LeCaRDv2 &
\citet{li2023LeCaRDv2LargeScaleChinese} &
Legal Relevance &
Characterization, Penalty, Procedure &
Retrieval Recall &
800 queries / 55K candidates &
Single-turn &
Rule-based &
LLM \\
\addlinespace
LawBench &
\citet{fei-etal-2024-lawbench} &
Legal Cognitive Ability &
Memorization, Understanding, Applying &
Accuracy, F1, ROUGE-L &
20 tasks &
Single-turn &
Rule-based (Regex) &
LLM \\
\addlinespace
LegalBench &
\citet{guhaLegalBenchCollaborativelyBuilt2023} &
Reasoning Breakdown &
Six reasoning types (IRAC) &
Accuracy; Balanced Acc &
162 tasks &
Single-turn &
Rule-based &
LLM \\
\bottomrule
\end{tabularx}
\end{minipage}
}
\vspace{-200pt}
\end{sidewaystable}

\end{document}